\pdfoutput=1
\documentclass[12pt]{spieman}  
\usepackage{amsmath,amsfonts,amssymb}
\usepackage{graphicx}
\usepackage{setspace}
\usepackage{tocloft}
\usepackage{xspace}
\usepackage{natbib}

\newcommand{\nirwals}{\textsc{nirwals}\xspace}

\title{The NIRWALS instrument detrending software pipeline}

\author[a]{Ralf Kotulla}
\author[a]{Marsha Wolf}
\author[a]{Matt Bershady}
\affil[a]{Department of Astronomy, University of Wisconsin-Madison, 475 N Charter St, Madison, WI 53706, USA}

\cftpagenumbersoff{figure}
\cftpagenumbersoff{table} 
\begin{document} 
\maketitle

\begin{abstract}
This document describes the basic instrument detrending software for the NIRWALS spectrograph on the SALT telescope. Its basic purpose is to process multiple non-destructive reads of increasing exposure time into a final image reflecting the observed source intensity as expressed in counts (or electrons) per second, including its uncertainty. The output products of this pipeline can then be used to as input for follow-up data processing to apply wavelength solutions and extract fluxes for individual fibers. All pipeline code is implemented in python and can be obtained via common portals such as PyPI and GitHub. Additional information on how to use the code is available online at nirwals.readthedocs.io.
\end{abstract}



\begin{spacing}{1}   

\section{Introduction}
The Near Infrared Washburn Astronomical Laboratories Spectrograph (NIRWALS; PI: Wolf, see \citealp{Wolf2022,Oppor2022,Smith2022} for technical details) is a near-infra\-red integral field spectrograph installed on the Southern African Large Telescope (SALT), extending SALT’s capabilities into the near infrared and providing medium resolution spectroscopy at R = 2000-6000 between 800 nm $< \lambda <$ 1700 nm using a Hawaii-2RG detector (also see \citealp{Mosby2016}). Similar to other NIR detectors this allows for non-destructive reads capable of sampling the integrated light as it accumulates during an exposure. On the flip side, this capability requires a more complex pre-processing before data can be used for spectral extraction. Here we describe the algorithm and implementation of the initial instrumental signature removal processing and present results for the detector performance and stability.

\section{Data reduction algorithm}

Data acquisition for \nirwals is a two-stage process. In the first step the detector is being read out by the actual detector controller; however, at this stage the available metadata is limited to only detector-related data, such as exposure settings, integration times, read number, etc, with no information about the configuration of either telescope or spectrograph. In a second step, a custom software gathers all auxiliary information, including but not limited to telescope pointing, environmental data, and spectrograph setup such as grating angles, focus, etc. and adds this information to the FITS header to be used during the data processing and for archiving purposes. The final raw data products supplied to the user thus includes significant metadata. All \nirwals filenames follow the naming convention N$<$date$><$sequence$>$.$<$sequence$>$.$<$ramp$>$.$<$group$>$.fits, where $<$date$>$ is the local date at the beginning of the observing night, $<$sequence$>$ is a running number reset daily, $<$ramp$>$ denotes the current ramp, and $<$group$>$ is the group number of each read. Using this convention we can construct the filename of each individual read frame in a given exposure sequence.

\subsection{Load read frames and mask saturated pixels}
As the first step we load a single frame from a given read sequence to extract relevant metadata, including the number of groups and reads and the detector setup including integration times and gain settings. The data from this "reference" frame is then enriched with information during the data processing and  propagated to the final output product. Using the \nirwals filename convention and the metadata from the reference frame the pipeline then loads each frame into memory so we have access to all reads for processing. After loading from disk, we flag each pixel with a read value exceeding a pre-determined saturation level (either a global or per-pixel value) as saturated; all pixels deemed to be saturated are excluded from all further analysis.  

\subsection{Reference pixel correction}

\begin{figure}
    \centering
    \includegraphics[width=\columnwidth]{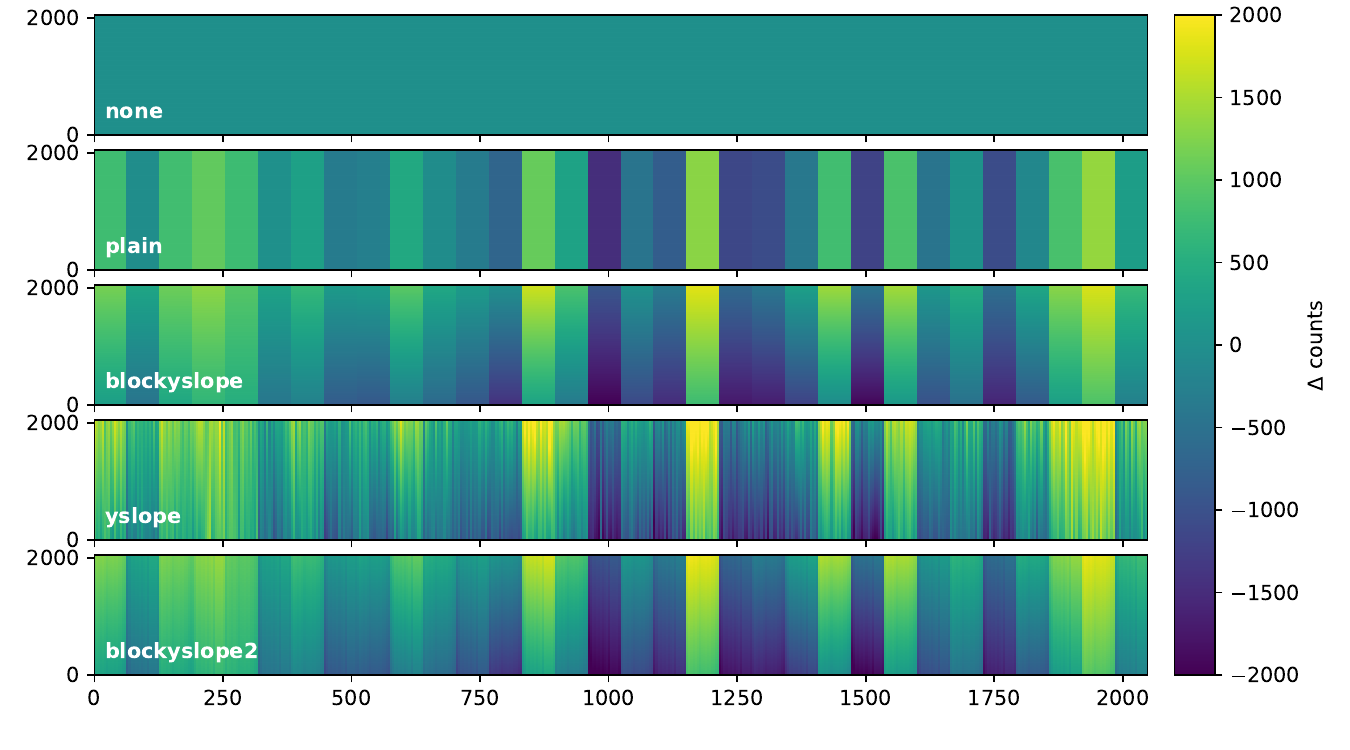}
    \caption{Comparison of different algorithm to generate a reference pixel correction; see text for details}
    \label{fig:refpixels}
\end{figure}

Similar to other NIR detectors (e.g. \citealp{Rauscher2004,Rauscher2007b}), our \nirwals detector comes with four rows and columns of "reference" pixels that are insensitive to light but are otherwise read out identical to all other pixels that can be used to correct for amplifier-specific bias levels. Our pipeline supports several different algorithms (see both panels in Figure \ref{fig:refpixels}) to use the data from these reference pixels and determine the bias level appropriate for each pixel across the full array. Figure \ref{fig:refpixels} illustrates both the actual correction for a representative \nirwals dark frame (top panels) and the resulting full-frame after correcting with the derived reference pixel map (bottom panels). In \emph{plain} mode we compute the median intensity level by combining the top and bottom 4 rows of each amplifier. compared to the frame without any correction the different background variations between amplifiers are already much reduced, but with a remaining vertical gradient. To address this, our \emph{yslope} and \emph{blockyslope} algorithm adds a linear interpolation between the median intensity observed in the top and bottom reference pixel areas; \emph{yslope} does so column by column, leading to a streaky appearance due to noise, whereas \emph{blockyslope} combines all reference pixels into a single value, thus decreasing its sensitivity to random fluctuations. The resulting correction is already much improved (see bottom left panel in Fig 1b), but we still observe a lower-amplitude horizontal pattern consistent across the read direction of all amplifiers (alternating amplifiers are read out in alternating directions; a linear ramp thus appears in the images as sawtooth pattern across two adjoining amplifiers). Our \emph{blockyslope2} algorithm averages this intra-amplifier pattern across all amplifiers  

The corresponding sub-panel in the bottom plot in Figure \ref{fig:refpixels} shows the dark frame after applying the reference pixel corrections generated with each of the four algorithms: \emph{none} is without correction, clearly showing the different amplifiers; using the \emph{plain} algorithm the amplifier differences are much reduced; but still showing a vertical gradient, which in turn is mitigated using the \emph{blockyslope} method; finally, the \emph{blockyslope2} algorithm also minimizes the sawtooth pattern across amplifiers.

\begin{figure}
    \centering
    \includegraphics[width=\columnwidth]{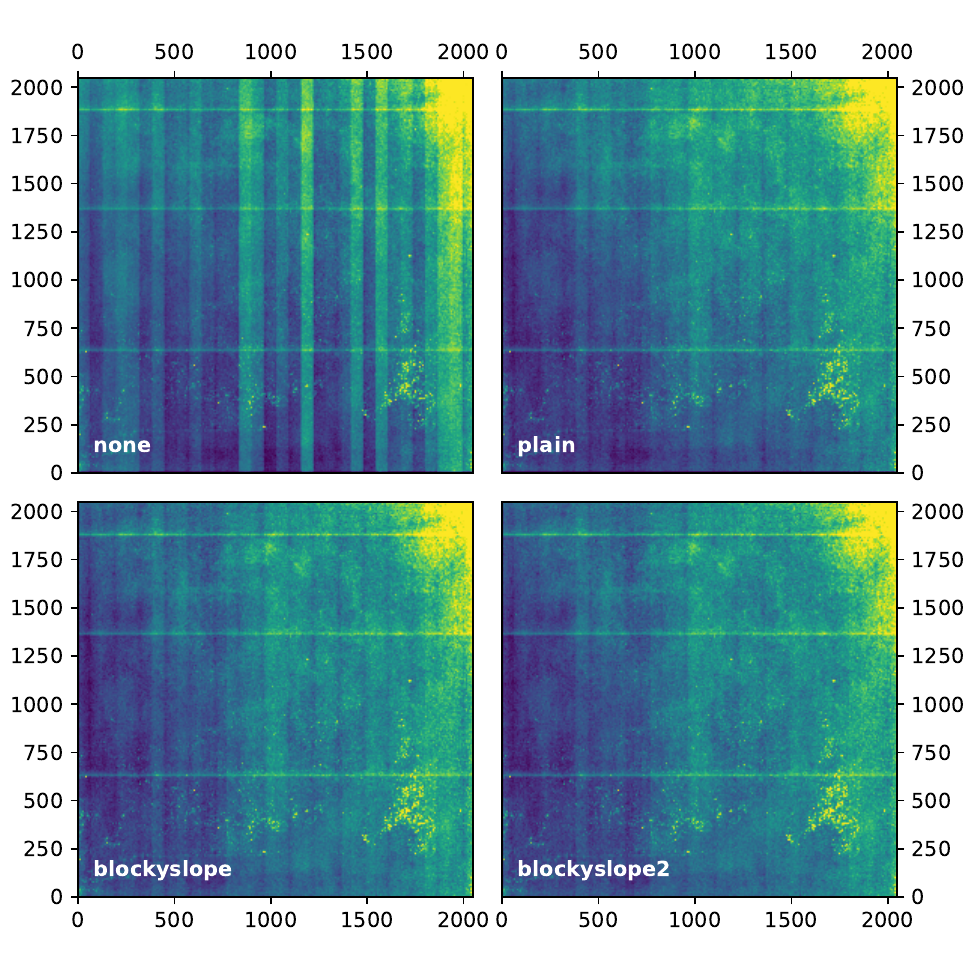}
    \caption{Example frame after subtracting reference pixel corrections derived using different algorithms; see text for details.}
    \label{fig:refpixels2}
\end{figure}

We also evaluated using the reference pixels on the left and right edges of each frame, but found no noticeable improvement from applying additional horizontal corrections.

In a final correction we evaluate the minimum corrected read value encountered in each pixel across the entire read sequence, and subtract this value from all reads. This ensures that the minimum value at the start of a sequence is approximately zero, and thus improves the non-linearity correction in the next step.

\subsection{Non-linearity correction}
\begin{figure}
    \centering
    \includegraphics[width=\columnwidth]{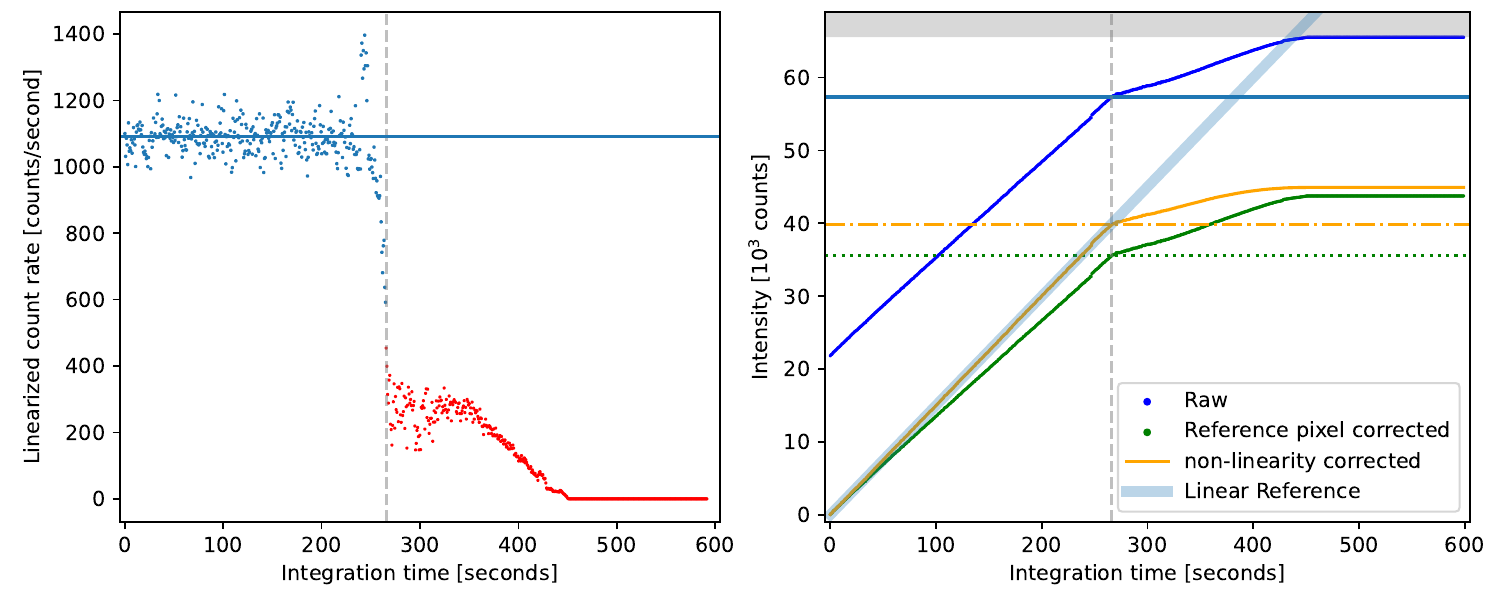}
    \caption{Illustration of the algorithm to determine the non-linearity fit for each pixel. \emph{Left panel:} Differential count-rate (i.e. difference in flux between subsequent reads, divided by incremental integration time), \emph{after} iteratively applying the latest fit parameters; The break denoted by the vertical dash lined marks the onset of detector saturation. \emph{Right panel}: Observed pixel intensitities before (blue data points) and after (green) subtracting the reference pixel correction; The wider blue line marks the ideal, linear intensity curve used as target reference; Shown in orange is the observed intensity \emph{after} applying the nonlinearity correction; horizontal lines marks the saturation limits (blue) as well as raw (green) and non-linearity corrected (orange) full-well depths.}
    \label{fig:nonlinearity}
\end{figure}

One of the most important steps in \nirwals data processing is to correct for the non-linearity of the sensitivity of individual pixels to incoming light, manifesting as a decrease in the slope of integrated count rate as function of time. To account for this we use integration sequences where we illuminate the detector with diffuse light (similar to a flat-field) of constant intensity to obtain the observed amplitude as function of time; by assuming a linear response at low intensities we can then convert from integration time to "true" intensity (as would be observed with a perfectly linear detector), and from there derive a polynomial fit of true intensity as function of observed intensity. In practice we choose a fifth-order polynomial to describe the fit, and treat every pixel in the array independently (see Figure \ref{fig:nonlinearity} for some examples for why this is required). Close to saturation level the observed relation between measured intensity and time flattens off significantly, and we also derive the point of this flattening off to determine the saturation level (and full well depth) for each individual pixel.

\begin{figure}
    \centering
    \includegraphics[width=\columnwidth]{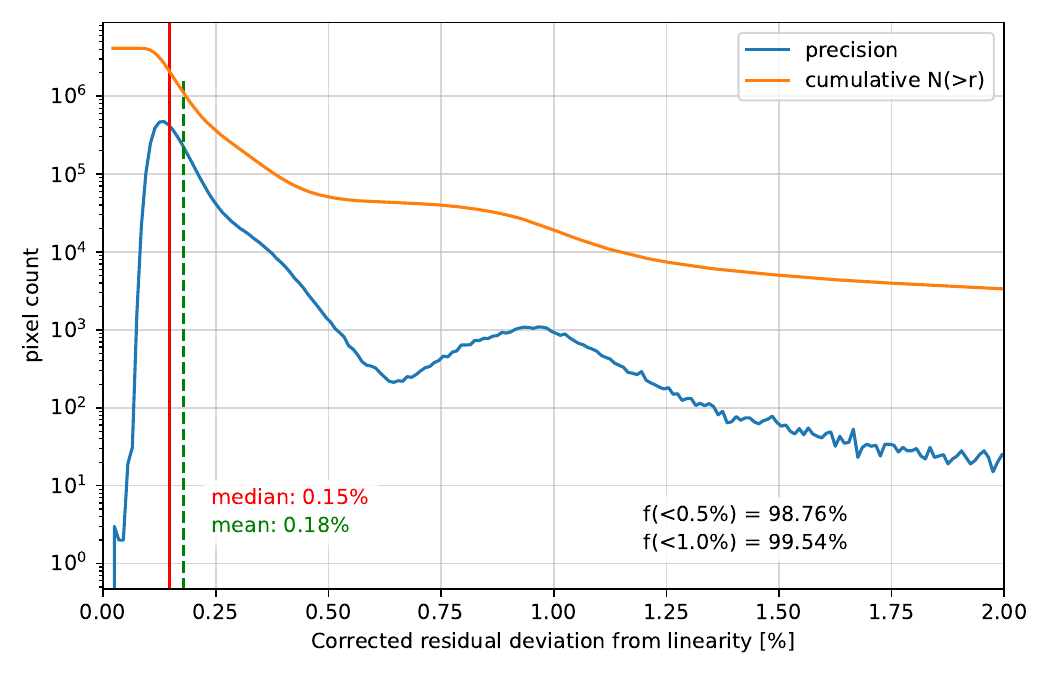}
    \caption{Residual nonlinearity \emph{after} correcting for nonlinearity using our polynomial fits. The blue curve shows the distribution of the fractional nonlinearity remaining after applying the best-fit solution to each pixel. Shown in orange is a cumulative distribution showing the number of pixels N($>$r) with residual nonlinearities larger than a given value r.}
    \label{fig:resid_nonlin}
\end{figure}

Using this polynomial fit we can now compute the nonlinearity corrected data and compare that to the "true" intensity used for fitting. A ratio between the corrected data and this true intensity then allows us to derive the quality of the correction. We therefore compute, for each read, the fractional uncorrected residual, and from the 1$\sigma$ width of its distribution across all reads derive a quality metric. Figure \ref{fig:resid_nonlin} shows the results across all pixels with valid nonlinearity fits. Using a 5th order polynomial we find a median residual nonlinearity across all pixels of 0.15\%, demonstrating a very good correction. Shown in the orange line in Fig. \ref{fig:resid_nonlin} is the cumulative distribution; While there is a long tail out to larger residuals, the total number of pixels in this tail is small: all but 0.5\% of all pixels are linearizable to better than 1\%, with 98.7\% of all pixels to better than 0.5\%.

Alongside the non-linearity correction polynomials we also determine the saturation level for each pixel by finding the level at which the corrected count rate (as expressed in counts per second) drop significantly from a constant level at lower observed intensities. The combined non-linearity fitting and determination of the saturation level is then iterated several times until both fit and saturation levels converge. A graphical representation of the method and results for a single pixel are shown in Figure \ref{fig:nonlinearity}. Note that in general the useful saturation level (i.e. the level where the observed level can no longer be linearized, at least not with a single polynomial fit) occurs significantly below the numerical saturation level of 65,535 counts (in the example shown at $\approx$ 57300 counts; see horizontal blue dashed line). 

During fitting we make note for which pixels this algorithm fails and for what reason (e.g. jumps in the data, a shape that is incompatible with our choice of polynomials, or data with an insufficient number of read samples as a result of high dark current). All error scenarios are handled and flagged so that these pixels can be excluded from any subsequent scientific interpretation.

\subsection{Combine up-the-ramp sequences into count rate images}
At this point in the reduction we not have a full data-cube, assembled from the reference-pixel and non-linearity corrected non-destructive reads sampling the integrated signal as a function of time ("up the ramp") for a given observing sequence. The next step is to process this data-cube into a single image representing the observed count rate. For this purpose we implement two different algorithms, with an additional pre-processing step outlined below. One algorithm follows the prescription from \cite{Rauscher2007}, computing the desired count rate as linear slope from the provided pairs of integration time and integrated flux. Alternatively we can perform a linear regression that also accounts for uncertainties in each observed read, and can be used to derive an uncertainty estimate for the derived count rate.

\begin{figure}
    \centering
    \includegraphics[width=\columnwidth]{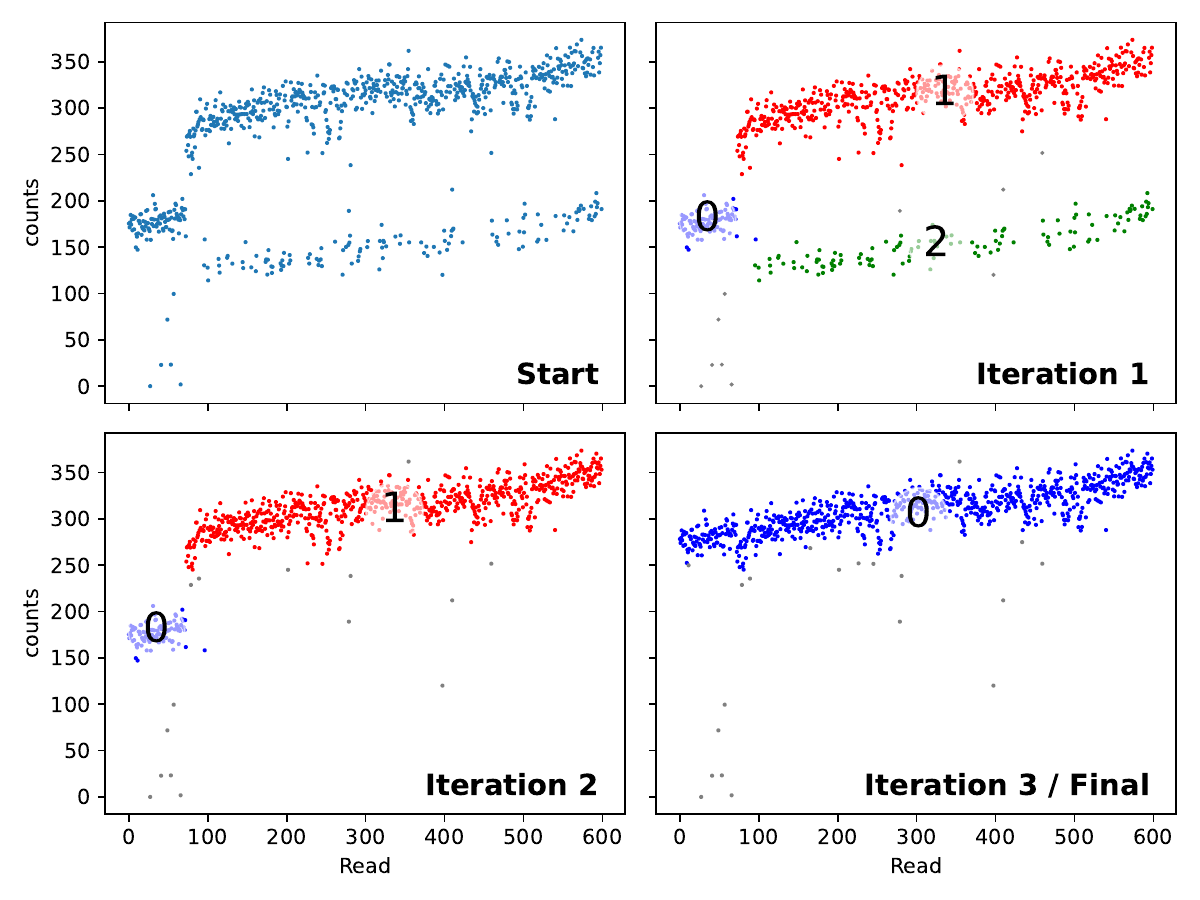}
    \caption{Step-by-step example of our recombine algorithm to correct for random telegraph noise; see text for details.}
    \label{fig:recombine}
\end{figure}

While these algorithms are computationally efficient they are not directly suitable to address two issues with the data: cosmic ray hits, as well as Random Telegraph Noise (RTN, see \citealp{Rauscher2007b}; alternative names for the same effect are "popcorn mesa noise" \citealp{Rauscher2004} or "burst noise" \citealp{Bacon2005}). An example for RTN is given in Figure \ref{fig:recombine}.

Figure \ref{fig:recombine} shows our implementation of an algorithm capable of mitigating or at least minimizing the effect of RTN. The prepared input data, i.e. observed counts as function of read, is shown in the top left panel; the effect of RTN is clearly visible as the sparser populated parallel sequence. Additionally, this sequence shows the impact of a cosmic ray hit around read \#100. To automatically identify the individual parts we apply the common unsupervised cluster algorithm DBSCAN to our read/flux data pairs. DBSCAN was chosen for its ability to reliably identify groups that may be overlapping when only considering read and flux values in isolation (e.g. when the RTN offset is less than the total accumulated signal) and its computational efficiency, especially when considering the need for iteration (see below) and the large number of pixels and reads in typical science integrations. 

In the present example this identifies 3 groups, shown with different colors in the top right panel of Fig. \ref{fig:recombine}. For each group we require a minimum number of "members" (based on the total number of reads). Individual data points that do not belong to any group are also identified and shown as grey data points; in our example these are several points with either larger noise or reduced RTN amplitude (between groups 1 and 2, as well as below group 0). 

In a next step we then identify the group with the most member data points, in this case group 1, and fit a linear slope to that data to yield a slope and intercept. If there are other groups that overlap this group temporally (as in the case of RTN), we can derive an average offset from the vertical distance of each point in group 2 to the average trend line derived for group 1, and bring both groups into alignment by applying this offset to the smaller group (here group 2). The results are shown in the bottom left panel of Fig \ref{fig:recombine}. After this step we go back and re-group data using the same algorithm. 

If there are no overlapping groups identified we also look for adjoining but separate groups, such as groups 1 and 0 shown in the lower left panel. A common reason for this offset are hits by cosmic rays, which add additional signal as a discrete read, leading to two parallel sequences offset in flux by the additional flux deposited by the cosmic. in this case we fit both sequences with individual slopes, yielding two sets of slopes and intercepts. We also identify the mid-point between both sequences (in the case of a cosmic the sequences are likely abutting, but this need not be the case in case additional RTN). Using the linear regressions for each sequence we can then derive a flux offset from the differences in extrapolated flux at this mid-point; the sequence with the fewer data points can then be flux-shifted by this offset to bring it into alignment with the other sequence.  

Lastly, this algorithm is repeated until either a pre-defined number of iterations has been attempted, or the clustering algorithm only identifies a single group (see bottom panel in Fig. \ref{fig:recombine}, potentially with some disparate outlier data points. 

Once all data has been "recombined" we can then progress to compute the best-fit count rate using the algorithms detailed earlier. Note that adding or subtracting offsets to different parts of the dataset affects the overall observed intensity for a given read and thus the final derived intercept point; however, the main scientific information is found in the observed \emph{slope}, and our algorithm is designed to not alter the slope of any given group of data points (although, if the matching of vertical offsets is imperfect, this may somewhat alter the derived slope, but in any case the so processed data is a significant improvement over deriving a slope from the initial uncorrected data).

A critical parameter for the DBSCAN algorithm is its sensitivity or aggressiveness in finding groups. If selected too large we loose sensivitity to small amplitude cosmics and/or RTN, while a too aggressive clustering may break up sequences into too many groups, leading to less reliable results and a dramatic increase in processing time. The presently chosen value appears to yield good overall results, but some future work tuning this sensitivity based on data properties (such as number of reads or estimated count rate) could be explored for even better performance. 

\section{Detector characterization}

\subsection{Bad pixel masking}
\begin{figure}
    \centering
    \includegraphics[width=\columnwidth]{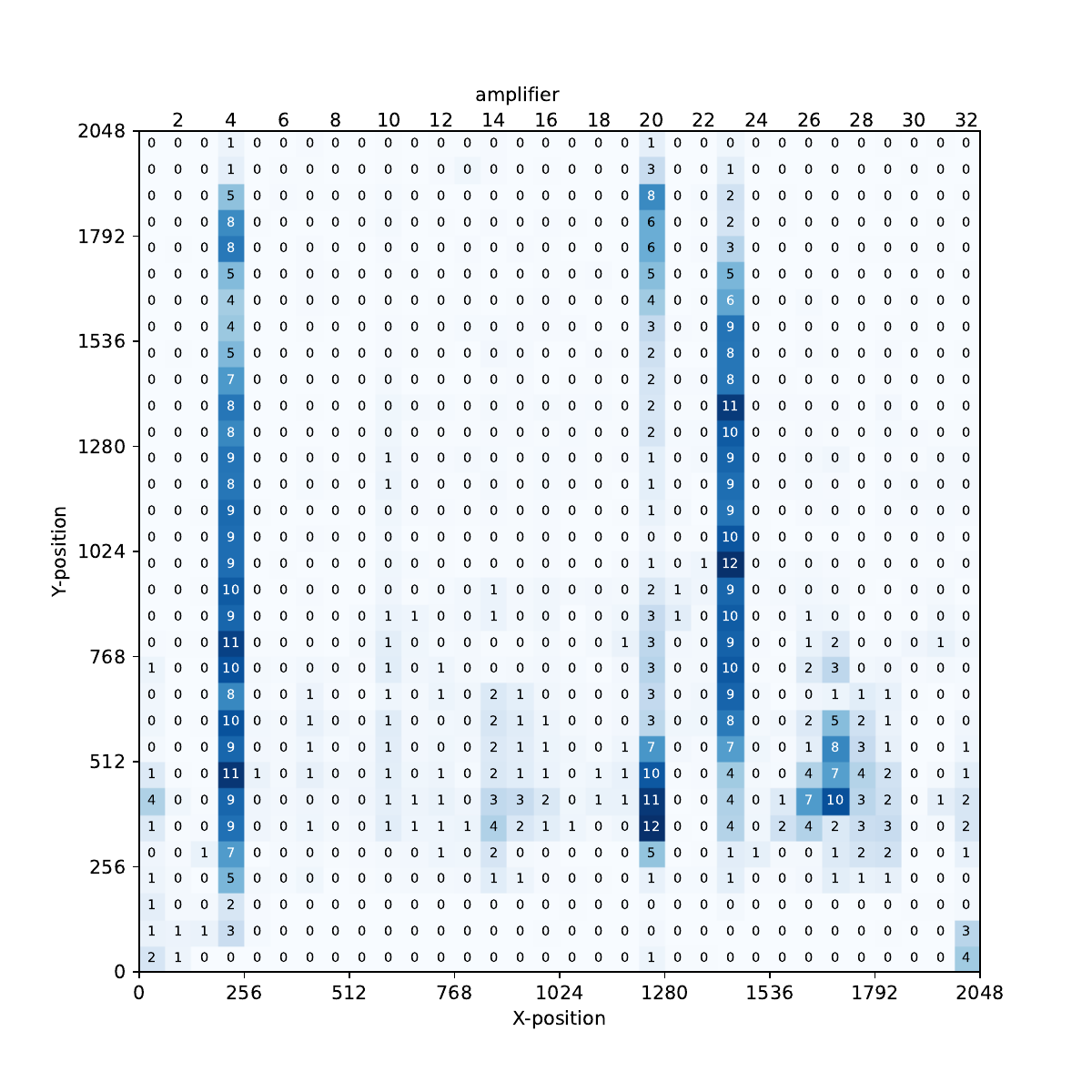}
    \caption{Fraction of pixels with problematic fits as determined during the nonlinearity correction fitting, in percent. Each block is 64x64 pixels in size. The three amplifiers with a higher fraction of bad pixels (amplifiers 4, 20, and 23) also correspond to the amplifiers with lower quality gain values (see Sect. \ref{sect:gains}).}
    \label{fig:badpixels}
\end{figure}

As mentioned before, during nonlinearity fitting we also keep track of why the fit failed for each of the pixels. Figure \ref{fig:badpixels} shows a graphical representation of the fraction of pixels with bad fits. Notably, most of the problematic pixels are constrained to a few areas, with concentrations in amplifiers 4, 20, and 23 (the same amplifiers with poorly constrained gain values), as well as an area towards the lower right (around coordinates x=400,y=1700).

Among the bad pixels, more than half failed due to insufficient data or because no idealized linear slope could be estimated, in both cases likely arising in hot pixels that saturate within only a few reads and thus do not provide sufficient sampling for a full fit. A second major category ($\sim$40\% of bad pixels) are made up of pixels for which the polynomial fit resulting in a negative linear term, i.e. with a very strong nonlinearity that no longer allows to reconstruct a physically plausible linearized solution.  

During data processing, all fitting flags are propagated from the nonlinearity coefficient file into the final data product, and can then be used to mask out problematic pixels during any further processing.

\subsection{Full well depth}

\begin{figure}
    \centering
    \includegraphics[width=0.49\columnwidth]{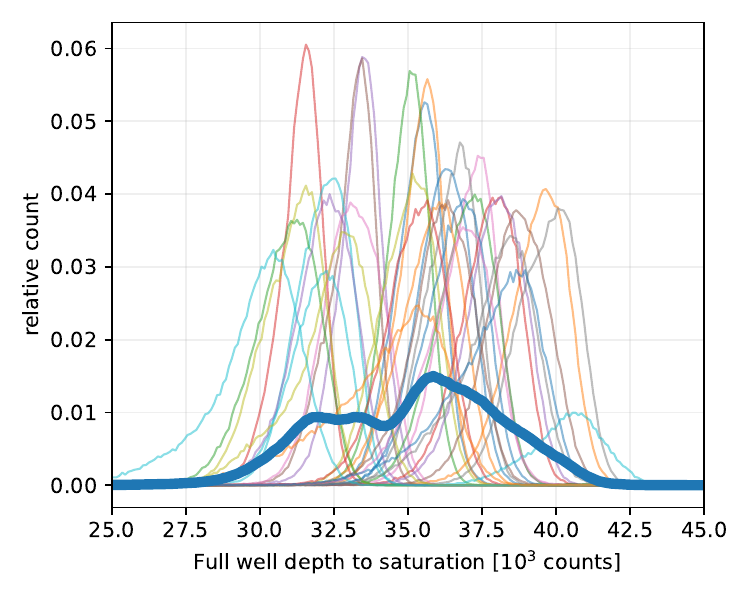}
    \includegraphics[width=0.49\columnwidth]{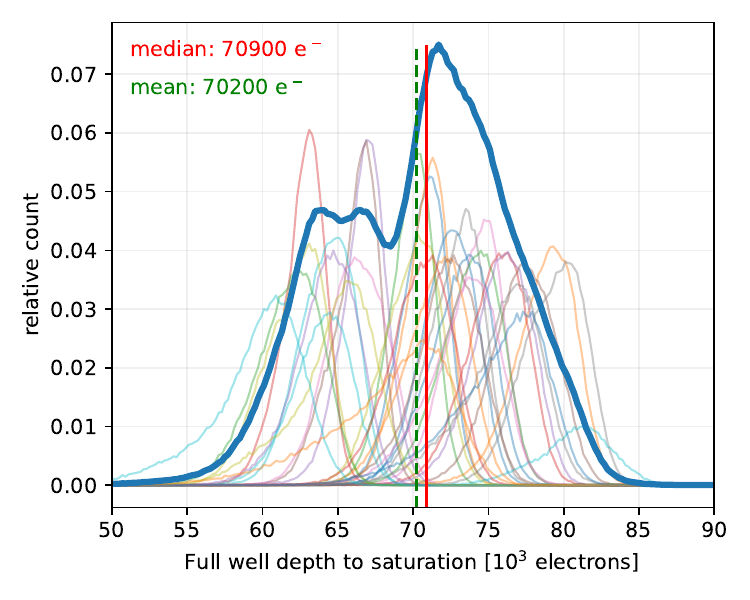}
    \caption{Observed full well depth to onset of uncorrectable non-linearity measured in counts (left panel) and electrons (right panel). Thin lines show results for individual amplifiers, solid blue lines show for the full detector.}
    \label{fig:fullwelldepth}
\end{figure}

Detector full well depth is indirectly determined as part of the nonlinearity fitting, in the form of the maximum uncorrected flux before the onset of non-linearity (see the intersection between the horizontal green line with the vertical grey line in Figure \ref{fig:nonlinearity}. Generally, each pixel has its own full well depth determined in this fashion, but we can combine results for all pixels in a given amplifier or the detector as a whole to derive an average full well depth. The results are shown in Figure \ref{fig:fullwelldepth}. Note that there are significant differences between different amplifiers.

\subsection{Data provenance}
A critical part of each generated data product is a complete inventory on how this product was generated, to ensure the data processing can be repeated and reproduced in the future. Towards this end, each generated output file contains a table with provenance metadata. This includes a full inventory of all files read during the processing (including all input files, non-linearity coefficients, etc), a log of all user-specified tunable parameters (e.g. what algorithm was used for URG fitting and/or reference pixel correction), as well as some information about the software setup in use (most importantly, what version of the pipeline was being used).

\subsection{Gain calculation}
\label{sect:gains}
To estimate the gain of our \nirwals detector we largely follow the differential photon transfer curve method detailed in \cite{Rest2017}. This method uses a variation of the photon transfer curve technique commonly used in conventional CCDs to derive a differential gain from pairs of reads in two identical flat-field and dark sequences to calculate the noise as 
\begin{equation}
g = \frac{\overline{\left((FF1a-FF1b)-(D1a-D1b)\right) + \left((FF2a-FF2b)-(D1a-D2a)\right) }}{(\sigma^2_{FF1a-FF1b}+\sigma^2_{FF2a-FF2b})-(\sigma^2_{D1a-D1b}+\sigma^2_{D2a-D2b})}
\end{equation}

where FF1 and FF2 (D1 \& D2) are two separate but otherwise identical flatfield (dark) sequences, and Xa and Xb are different read frames from each sequence. For each read we can then derive a differential gain measurement; to generate an effective gain we fit all gain values with a straight line and find the extrapolated gain at zero intensity.

\begin{figure}
    \centering
    \includegraphics[width=\columnwidth]{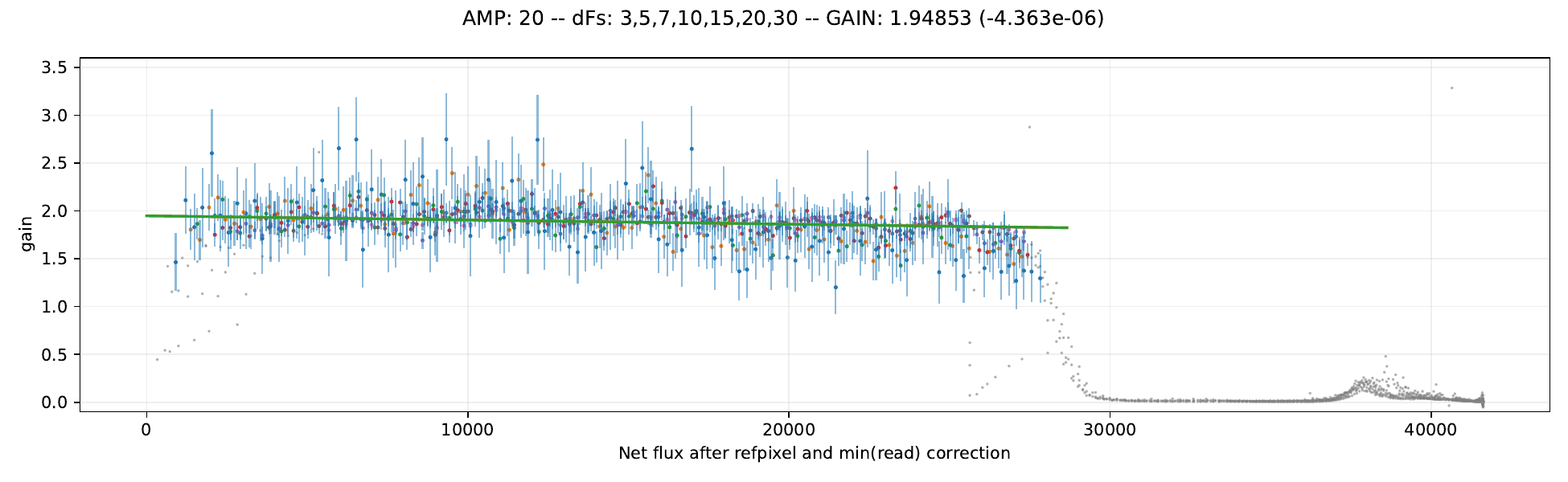}
    \includegraphics[width=\columnwidth]{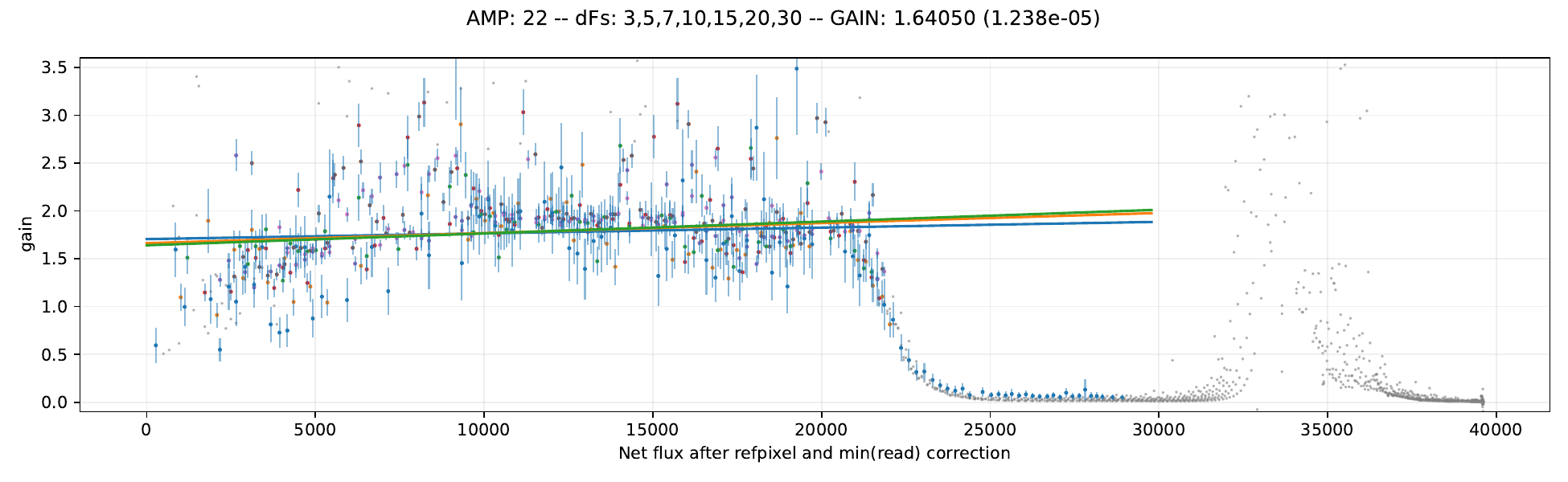}
    \caption{Example for results from our enhanced differential photon transfer curve method for two amplifiers. The top panel shows amplifier 20, for which all measurements yield a reliable fit to gain as function of intensity, with the expected slightly negative slope due to detector nonlinearity. The bottom panel shows amplifier 22, where unstable reads, especially at integration intensities below $10^4$ counts skew the results and cause problems for accurate gain estimation.}
    \label{fig:gains}
\end{figure}

In an ideal world we would be able to use subsequent reads to finely sample this gain versus read or intensity curve. However, due to cosmetic and other issues with this particular detector this is unfortunately not as easy. First, subsequent reads in our flatfield sequences show too little signal to yield reliable results, so instead of subsequent reads to use reads spaced dF reads apart (for example, instead of using reads 4-3 we use reads 10-3) to increase signal. Larger dF values increase signal (the numinator in eq. X), but at the expense of larger systematic variations due to detector read issues. Additionally, due to the poor detector cosmetics we can not use the entire amplifier to derive the $\sigma X$ noise values; to address this we a) manually masked out areas across the detector with an above average number of bad pixels, and b) added a comprehensive outlier clipping to our algorithm. To select pixels to include in the calculation we iteratively sigma-clip outliers in each of the 8 flatfield and dark reads (FF1a,FF1b,FF2a etc), as well as in the difference images (FF1a-FF1b, etc), keeping only pixels not marked as outliers in any of these 12 (difference-) images. Lastly, we sample the data repeatedly, using a range of dF values ranging from 3 to 30 for best results. This method successfully yields reliable and reproducible gain values for most but not all amplifiers; Fig. \ref{fig:gains} shows examples for a good and a less reliable amplifier. 

\section{Remaining issues}

While the actual data processing is settled on a working algorithm, there remain a few issue, mostly dealing with crosstalk, that could possibly improved in the future as the detector system is better understood.

\subsection{Correction of crosstalk across amplifiers}
Detector testing during commissioning revealed a low-level crosstalk between amplifiers, where a high signal in one amplifier (e.g. due to a hot pixel, cosmic ray, or sky emission line) causes a slight negative depression in all other amplifiers at location being read out simultaneously. Given the complexities of 32 amplifiers (each serving as a simultaneous source and sink of this effect) no such crosstalk correction has been implemented into the pipeline.

\subsection{Adjacent pixel crosstalk}
\begin{figure}
    \centering
    \includegraphics[width=\columnwidth]{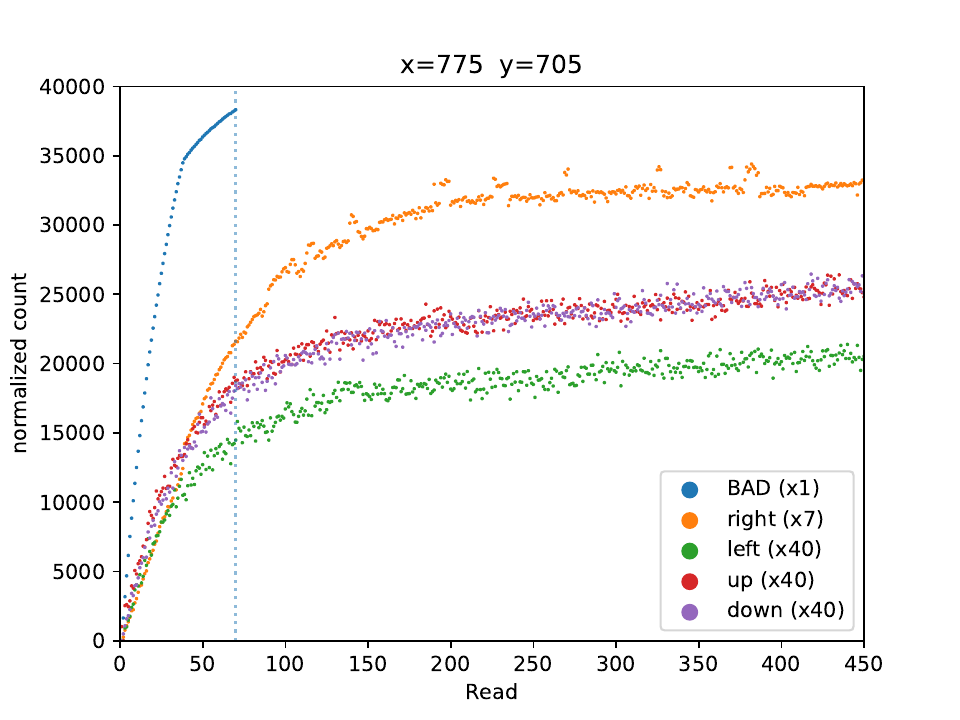}
    \caption{Example of next-door pixel crosstalk in a dark frame. Shown in blue is a central hot pixel that quickly saturates; Also shown are the four adjacent pixels (left, right, up \& down). Note the different flux scaling factors for each of the pixels.}
    \label{fig:nextdoorcrosstalk}
\end{figure}

Another source of crosstalk was found more locally, with pixels having an effect on their direct neighboring pixels. An example is also illustrated in Figure \ref{fig:nextdoorcrosstalk}, for a given hot pixel in a dark exposure sequence. 

While the central pixel is accumulating signal, it seems to have a direct impact on its 4 neighboring pixels, in particular on the pixel read out right after the hot pixel (pixel 'right' in this example; depending on amplifier and thus readout direction this may als o be the pixel to the left). Once the hot pixel has saturated and thus no longer accumulates additional signal this effect is greatly diminished, and the neighboring pixels continue to accrue signal at a much lower -- and likely more realistic -- rate until the end of the exposure. 

As a result, each hot pixel thus has the potential to also negatively impact its four neighboring pixels, and thus further degrades the cosmetics of the detector. While this effect could possibly be addressed or mitigated, e.g. by masking out the initial reads of the four affected pixels read out \emph{before} the hot pixel saturated, or even by correcting those initial reads based on the accumulated signal or signal rate of the hot pixel, more work would be required to study this effect in more detail.

\section{Implementation}

All functionality is implemented in python, and available as \texttt{nirwals} package using the regular distribution channels such as GitHub and PyPI. 

For user convenience the nirwals package also includes a number of standalone scripts that execute the python code. These include:
\begin{itemize}
    \item \texttt{nirwals\_reduce} for the actual reduction
    \item \texttt{nirwals\_fit\_nonlinearity} to fit non-linearity polynomials from a flat-field exposure ramp. 
    \item \texttt{nirwals\_provenance} to read the provenance information from the corresponging table embedded into each output file.
\end{itemize}

Documentation is available online at https://nirwals.readthedocs.io/.





    





\bibliography{references}   
\bibliographystyle{aasjournal}   








\end{spacing}
\end{document}